\documentclass[aps,showpacs,preprintnumbers,amsmath,amssymb,
twocolumn,
eqsecnum
]{revtex4}

\usepackage{amsmath}

\input epsf

\begin{document}

    \bibliographystyle{apsrev}

    \title {VECTOR BOSONS IN STRONG FIELDS}
                
    \author{M.Yu.Kuchiev}
                \email[email:]{kmy@phys.unsw.edu.au}
    \author{V.V.Flambaum}
                \email[email:]{flambaum@phys.unsw.edu.au}
    \affiliation{School of Physics, University of New South Wales,
      Sydney 2052, Australia}
    \date{\today}

\begin{abstract}
The Coulomb  problem for charged massive vector bosons is known to be unstable,
 the boson falls on the Coulomb center. It is shown that when the charge of the
 Coulomb center is smeared over a small but finite volume, then instead of the
 fall there appears a large series of bound states localized inside the volume
 containing the Coulomb charge. It is shown also that the fall on the Coulomb
 center is completely suppressed, when the conventional QED vacuum polarization
 is taken into account.
 In gauge theories (SU(2), Standard Model) the
 renormalizability of a theory may be insufficient to guarantee an absence of
 the wave function collapse. Another interesting feature:
  the density of charge of vector bosons can have a ``wrong'' sign.
\end{abstract}

\maketitle

    \section{Introduction}
    \label{intro}
    Consider the Coulomb problem for charged, massive, vector bosons.
    It was known for a long time that Proca's formulation of the
    theory for massive vector particles \cite{proca_1936} produces
    inadequate results for the Coulomb problem
    \cite{massey-corben_1939,oppenheimer-snyder-serber_1940,tamm_1940-1-2}.
    This fact inspired Corben and Schwinger
    \cite{corben-schwinger_1940} to reformulate the theory, tuning
    the Lagrangian and equations of motion in such a way as to force
    the hyromagnetic ratio of the vector boson to acquire a favorable
    value $g=2$.  Later on the formalism of
    \cite{corben-schwinger_1940} was found to have a connection with
    the non-Abelian gauge theory \cite{schwinger_1964}, which makes it
    relevant for the present day studies.  
    Ref. \cite{corben-schwinger_1940} found a realistic discrete energy
    spectrum for the Coulomb problem for vector bosons, but it
    discovered also a fundamental flaw. Some series of
    quantum states (those with $j=0$ and so called  ``$l=0$'' states
    ) exhibited the fall of the
    boson on the Coulomb center. 
 The present work describes the remedy 
    that ``saves'' the Coulomb problem and several relevant phenomena
    \cite{flambaum-kuchiev-prl-07,kuchiev-flambaum-prd-06,kuchiev-flambaum-mpl-06,flambaum-kuchiev-mpl-07}:
 an existence of an additional infinite series of energy levels for the
 vector boson when the charge of the Coulomb center is smeared over a finite
 volume, the fact that the charge density of the positive vector boson can
be negative in some areas of space, relation between the fall to the
 center and sign of the Gell-Man - Low $\beta$-function.


    Consider boson fields in the electroweak part of the Lagrangian of
    the Standard Model, see e.g.  \cite{weinberg_2001},
   \begin{eqnarray}
      \label{gauge}
      {\mathcal L}= -\frac{1}{4}\,
      \left(\partial_\mu \boldsymbol{A}_\nu-\partial_\nu \boldsymbol{A}_\mu  +
        g \,\boldsymbol{A}_\mu \times \boldsymbol{A}_\nu\right)^2
      \\ \nonumber
            -\frac{1}{4}\,
      \left(\partial_\mu { B}_\nu-\partial_\nu {B}_\mu  \right)^2+
      \frac{1}{2}\,D_\mu\Phi^+ D^\mu \Phi.\quad
    \end{eqnarray}
    Here $\boldsymbol{A}_\mu$ and $B_\mu$ are the triplet of $SU(2)$
    and the $U(1)$ gauge potentials respectively (abridged notation is
    used here).  The covariant derivative $D_\mu\Phi$ takes into
    account that the Higgs field $\Phi$ has a hypercharge $Y=2$, which
    describes its interaction with the $U(1)$ field, and is
    transformed as a doublet under the $SU(2)$ gauge transformations. Taking the unitary gauge 
    one can present it via one real component
    $  \Phi^T =( 0, \phi)$,  $\phi\,*=\phi$.  
    Assuming that the scalar field develops the vacuum expectation
    value $\phi=\phi_0$ and the Higgs mechanism takes place, one finds
    that the gauge field can be presented as an electromagnetic field
    $A_\mu =-\sin \theta \,A_\mu^3+\cos\theta \,B_\mu$
    and a triplet of massive fields $W^\pm_\mu, \,Z_\mu$, where
$      W^-_\mu =\left(A_\mu^1-iA_\mu^2 \right)/\sqrt 2
		\equiv W_\mu$ is the $W$-boson with charge
    $e=-|e|$.  
    Expanding the Lagrangian (\ref{gauge}) in the vicinity of
    $\phi=\phi_0$ up to bilinear in the fields
    $W_\mu,W_\mu^+$ terms, one derives an effective Lagrangian
    \begin{eqnarray}
      {\mathcal L}^W \!= \!-\frac{1}{2}\left(\nabla_\mu W_\nu -
        \nabla_\nu W_\mu\right)^+ \!\left(\nabla^\mu W^\nu-
        \nabla^\nu W^\mu\right)  
        \\ \nonumber
        + i e \, F^{\mu\nu} W^+_\mu W_\nu + 
        m^2 W_\mu^+ W^\mu,~
         \label{W}      
    \end{eqnarray}
    which describes the propagation of $W$-bosons with the mass $m$ 
    in an external
    electromagnetic field, which appears in the derivative
    $\nabla_\mu=\partial_\mu +i e A_\mu$ and in the
    field $F^{\mu\nu}=\partial^\mu A^\nu-\partial^\nu A^\mu$. The
    first and the last terms in Eq.(\ref{W}) are present in the Proca
    formalism \cite{proca_1936}, the second one was introduced
    by Corben and Schwinger \cite{corben-schwinger_1940}.
    From Eq.(\ref{W}) one derives the equation of
    motion for vector bosons suggested in
     \cite{corben-schwinger_1940}
    \begin{eqnarray}
      \label{wave}
      \left( \nabla^2+m^2\right) W^\mu
      + 2 i e \,F^{\mu\nu}\,W_\nu-  
      \nabla^\mu \nabla^\nu \,W_\nu =0~.
    \end{eqnarray}
    Taking a covariant derivative here one finds
    $ m^2\nabla_\mu W^\mu+ie\,j_\mu W^\mu=0~$,	
	\label{jext}
    where $j^\mu=\partial_\nu F^{\nu\mu}$ is the external current. Its explicit 
    presence in the equations of motion
    distinguishes the case of vector bosons
    from the Dirac and scalar particles.
    The coefficient 2 in
    the second term ensures that the g-factor of the boson takes the
    value $g=2$.   
    From Eq.(\ref{W}) one derives the current of $W$-bosons
    \begin{eqnarray}
      j_{\phantom{\,}\mu}^{W}=-ie\Big( \,W^+_\nu \nabla_\mu W^\nu + 
      2 \nabla_\nu W^+_\mu  W^\nu -c.c. \,\Big) 
      \\ \nonumber
     -\frac{e^2}{m^2}(\, W_\mu^{+} W_\nu+W_\nu^{+}W_\mu\,)\,j^\nu~.
     \label{jW}
    \end{eqnarray}

    \section{Coulomb problem and Corben-Schwinger anomaly}
    To study an anomaly in the Coulomb problem for a $W$-boson, which was found in 
 		\cite{massey-corben_1939,oppenheimer-snyder-serber_1940,%
 		tamm_1940-1-2}and studied in detail in \cite{corben-schwinger_1940}
		let us presume that the $W$-boson propagates in the Coulomb field produced by 
    a heavy, point-like attractive center with the charge $Z|e|$, and neglect the width of the 
    $W$-boson.  In  
    \cite{corben-schwinger_1940} it was shown that the energy spectrum of this problem is 
    given by the Sommerfeld formula
\begin{equation}
\varepsilon\,=\,m\left(1+(Z\alpha)^2/(\gamma+n-j-1/2)^{2}\right)^{-1/2}~,
\label{som}
\end{equation}
where $\gamma\,=\,[ \,( j+1/2)^2-(Z\alpha)^2 \,]^{1/2}$, and $j=0,1\dots$ is the total momentum.
    
    However, the charge distribution of the $W$-boson in the $j=0$ wave (and also in another 
    wave, called ``$l=0$'') is so singular at the origin that the 
    boson falls on the Coulomb center. For the stationary state 
    of the boson its wave function can be expressed via the vector 
    $\mathbf{W}=\mathbf{W}(\mathbf{r})$, 
    which in the $j=0$ case reads $\mathbf{W}=\mathbf{n}\,v(r)$, where $\mathbf{n}=
    \mathbf{r}/r$. Eq.(\ref{wave}) in this case reads gives for $v$
    \begin{eqnarray}
       \big((\varepsilon-U)^2\!-m^2\big)v&=&
     -(\varepsilon-U)\frac{d}{dr} \left(\,
    \frac{v'+2v/r }{\varepsilon-U-\Upsilon}
       \,\right)\quad
       \label{CSj=0}
       \\ \nonumber
    &&-U'\,\frac{v'+2v/r}{\varepsilon-U-\Upsilon}.
    \end{eqnarray}
    Here $U$ is the potential energy, which for the Coulomb problem is $U=-Z\alpha/r$, and 
    $\Upsilon$ is an effective addition to the potential, 
 		\begin{eqnarray}
      \label{Ups}
      \Upsilon\,=\,e\rho/ m^2\,=\,-\Delta U/m^2~,
    \end{eqnarray}   
		which arises from the density of the external charge $\rho$ that is present in the current $j^\mu$ in (\ref{jext}).
		 For the purely Coulomb problem  
		$\Upsilon=0$ when 
		$r>0$, but the term $\Upsilon$ plays a very important role for more sophisticated cases 
		discussed below. From Eq.(\ref{CSj=0}) one derives $v\propto r^{\gamma-3/2}$, when 
		$r\rightarrow 0$. The density of the $W$-boson $\rho^W$ found from this estimate and 
		Eq.(\ref{jW}) is 
		\begin{eqnarray}
      \label{cdvb}
      \rho^W\,\propto \,\,-e\,r^{2\gamma-4}~,\quad\quad r\rightarrow 0~,
      \label{-}
    \end{eqnarray}
For $j=0$ one has $\gamma<1/2$, and Eq.(\ref{cdvb}) signals a nonintegrable divergence for the charge localized in a small vicinity of the origin $\int \rho^Wd^3r=\infty$, which indicates the fall of the boson on the Coulomb center, as was found in \cite{corben-schwinger_1940}, see also \cite{kuchiev-flambaum-prd-06}.

\section{Finite size of Coulomb center}
\label{finite}
When the Sommerfeld spectrum was obtained in
 \cite{corben-schwinger_1940}, the potential $\Upsilon$ (\ref{Ups})
was neglected since it is proportional to $\delta$-function. Below we show
that this potential destroys the discrete spectrum.
 Consider what happens when the
charge of the Coulomb center is smeared over some small, but finite volume. 
Take a model of the homogeneous distribution of the charge of the Coulomb center over the sphere with a small radius $R$, which satisfies	$Z\alpha\,\gg\, mR$.
Then the potential energy $U$ and the effective potential $\Upsilon$ (\ref{Ups})for  $r\le R$ are
\begin{align}
&U\,=\,\frac{Z\alpha}{2R}\left(\frac{r^2}{R^2}-3\right)~,
\label{U}
\\
&\Upsilon\,=\,\frac{e\rho}{m^2}\,=\,-\frac{\Delta\,U}{m^2}\,=\,-3\frac{Z\alpha}{m^2\,R^3}~.
\label{UpsR3}
\end{align}
We want eventually compress nucleus to a point, $R\rightarrow 0$.
In this case $ |\Upsilon|\gg|U|\gg \varepsilon\sim m$.
 Neglecting $\varepsilon,m$ and $|U|$ wherever possible in Eq.(\ref{CSj=0})we rewrite it inside the charged sphere in a form
\begin{equation}
	U\,v\,=\,- \frac{d}{dr}\left(\,\frac{v'+2v/r}{\Upsilon}\right).
	\label{form}
\end{equation}
With sufficient for us accuracy we approximate $U,\Upsilon$ by their values at the origin
$U\simeq -3 Z\alpha/(2R)~$,$\quad\quad\Upsilon \simeq -3 Z\alpha/(m^2R^3)~$,
rewriting after that Eq.(\ref{form}) as follows
\begin{align}
& k^2\,v\,=\,
\,- v''-\frac{2}{r}\,v'+\frac{2}{r^2}\,v~,
\label{k^2}
\\
& k\,=\,\frac{3}{ \sqrt{2} }\,\frac{ Z\alpha }{ mR^2 }~.
	\label{k}
\end{align}
Clearly Eq.(\ref{k^2}) describes a free motion of a particle with 
the momentum $k$ and the orbital momentum $l=1$, $v\propto j_1(kr)$, $j_1(x)$ is the spherical Bessel function.
Within the radius of the charged sphere the function $v=v(r)$ exhibits large number of oscillations
\begin{equation}
	N\,\simeq\,kR \,=\,\frac{3}{ \sqrt{2} }\,\frac{ Z\alpha }{ m\,R }\,\gg\,1~.
	\label{N}
\end{equation}
These oscillations indicate that there is a large number $N$ of bound states for the boson with the momentum $j=0$, the infinite number for $R\rightarrow 0$ . They are localized inside the charged sphere, and their energies
tend to minus infinity for $R\rightarrow 0$. This is a signature of the
absence of the  discrete spectrum ( vacuum breakdown, pair creation, etc). 
 The effective potential $\Upsilon$, which is specific for vector bosons, plays a key role in this phenomenon.

\section{Vacuum polarization}
\label{vacpol}
After discussing different perspectives on the problem of the fall of the boson on the Coulomb center, let us show that there is a simple mechanism that prevents this fall altogether.
Consider the conventional QED vacuum polarization, which manifests itself as an additional 
term in the potential energy of the the $W$-boson that reads
      \begin{align}
&     U(r)\,=\,-[\,1+S(r)\,] Z\alpha/r~,
\label{US}
\\   
&     S(r)\, \simeq \, -\alpha \beta \,\ln\left(m_{Z}r\right)~,\quad\quad r\rightarrow 0
\label{S}
      \end{align}
The term proportional to $S(r)$ in Eq.(\ref{S}) describes the vacuum polarization in the lowest order of the perturbation theory, when it is given by the Uehling potential. 
Eq.(\ref{S}) relies on an asymptotic behaviour of this potential. The logarithmic function there is related to the famous logarithm responsible for the scaling of the QED coupling constant $\alpha^{-1}(\mu)=\alpha^{-1}(\mu_0)-\beta\,\ln(\mu/\mu_0)$; the coefficient $\beta$ in the latter formula and in Eq.(\ref{S}) is same. This coefficient can be estimated from experimental data for the fine structure constant taken at the masses of the Z-boson and $\tau$-lepton, which gives $\beta \simeq 1.42$. This value complies with the simplest, ``naive'' estimate based on the quark model, which neglects the QCD vacuum and gives $\beta_\mathrm{est} \approx (2/3\pi)\sum_i (q^2_i/e^2)=1.70$, where $q_i$ is the charge of an
$i$-th quark. 

From Eqs. (\ref{Ups}),(\ref{US}) and (\ref{S})
      one derives for the $\Upsilon$-term
      \begin{eqnarray}
        \label{estUps}
        \Upsilon\simeq Z\alpha^2 \beta/ (m^2r^3),\quad
        r\rightarrow 0~,
        \label{pot}
      \end{eqnarray}
which proves to be large at the origin, being there the largest term in Eq.(\ref{CSj=0}) at small distances, $\Upsilon\gg |U|\gg \varepsilon,m$, $r\rightarrow 0$.  It is to be noted that the direct contribution of the vacuum polarization to the potential energy given by the term $S(r)$ in Eq.(\ref{pot}) is not pronounced, and is attractive, while the $\Upsilon$-term is large and have a positive sign. The dominant nature of the $\Upsilon$ term makes the problem under discussion similar to the one addressed in Section \ref{finite}. This fact makes Eq.(\ref{form}) derived there to be applicable in the context of the present discussion as well, though the $\Upsilon$-term this time is specified by Eq.(\ref{estUps}). With sufficient for our purposes accuracy we can rewrite the wave equation for the $j=0$ mode (\ref{form})
as follows
\begin{equation}
	0\,=\,-v''-\Upsilon U v = \left(-\frac{d^2}{dr^2}+\frac{Z^2\alpha^3\beta}{m^2 r^4}\right)\,v~.
	\label{v''YUv}
\end{equation}
Here Eqs.(\ref{US}),(\ref{estUps}) are taken into account. Clearly, the coefficient in front of $v$ in the right-hand side of (\ref{v''YUv}) describes an effective strong repulsive potential, $U_\mathrm{eff}=-\Upsilon U= Z^2\alpha^3\beta/(m^2 r^4)$, which drastically reduces the function $v$ at small distances. This reduction is described by the following solution of Eq.(\ref{v''YUv}) 
\begin{equation}
	v\,\simeq\,r^{-2} \,\exp[\,-\,Z \alpha (\alpha\beta)^{1/2}/(mr)\,]~.
	\label{exp-1/r}
\end{equation}
The derivation presented reproduces only the main, exponential factor here. In order to justify the preexponential one, it is necessary to consider those terms, which were neglected when Eq.(\ref{form}) was approximated by simple Eq.(\ref{v''YUv}). For these and other details see  \cite{kuchiev-flambaum-prd-06}, where Eq.(\ref{exp-1/r}) was firstly derived.

Eq.(\ref{exp-1/r}) shows that $v$ is exponentially suppressed at the origin, same it obviously true for the charge density $\rho^W\propto v^2$. This is in contrast with the pure Coulomb problem, when the density is singular at the origin, see (\ref{cdvb}), and the $W$-boson falls on the Coulomb center. Thus the vacuum polarization eliminates the fall of the W-boson on the center, making the Coulomb problem well defined. Conventionally, for the scalar and Dirac particles, the vacuum polarization provides a weak attraction. In contrast, for vector bosons, the vacuum polarization produces a strong repulsive effect at small distances. The $\Upsilon$-term plays a key role in this phenomenon.

  Note that in the asymptotically free theories $\beta <0$ and the
 vacuum polarization does not prevent the fall to the center. However,
 additional diagrams which appear  in these theories
(e.g. Compton scattering of $W$-boson) may eliminate infinite number
of states inside the Coulomb center which were considered in section 3.
The Coulomb problem in gauge theories SU(2) and the Standard Model
 SU(2)$\times$U(1) has been considered in our work 
 \cite{flambaum-kuchiev-mpl-07}. To eliminate collapse of the $W$ wave function
 to the center in these
theories one may need additional generations of fermions or scalars which
 change the sign of $\beta$-function to  $\beta >0$ at small distances.
  
\section{Density of charge can have ``wrong'' sign}
Eq.(\ref{cdvb}) indicates that in the vicinity of the Coulomb center the charge density of the
W-boson has the wrong sign. To see the point clearly consider the W$^-$-boson, which has the negative charge $e=-|e|$. Eq.(\ref{cdvb}) tells us that near the origin the charge density is positive.
Moreover, even for free motion, in the case of the standing wave
$\boldsymbol{W}=\boldsymbol{C}\sin(\boldsymbol{p}\cdot\boldsymbol{r})$ and the longitudinal polarization,
 $|\,\boldsymbol{C}\cdot\boldsymbol{p}\,|=|\,C\,|p$, the density 
\begin{eqnarray}
\label{standl}
\rho\!= \!\frac{2e|C|^2\!}{\varepsilon}\Big((g\!-\!1)p^2\!
  -\!\big((2g-1)p^2 \! - \!\varepsilon^2\big)
\sin^2 (\boldsymbol{p}\!\cdot\!\boldsymbol{r})\Big).
\end{eqnarray}
has no definite sign. We see that for $g>1$ and energy  $\varepsilon >m \sqrt{g/(g-1)}$ the charge density
 may take negative values. In the case of the $W$-boson,  $g=2$,
 the change of sign appears
for the energies $\varepsilon > \sqrt{2} m$.
 The minimum of density corresponds to
 $\sin^2 (\boldsymbol{p}\cdot\boldsymbol{r})=1$.
  In the Proca case $g=1$ the sign of the charge density is fixed.
However, for $g<1$ the charge density may be negative again.
 The minimum of the density in this case corresponds to
 $\sin (\boldsymbol{p}\cdot\boldsymbol{r})=0$. As we will show
below this behavior is explained by the contribution of the
electric quadrupole moment  of the vector particle, $Q \propto (g-1)e/m^2$.

To find a reason for this unusual behaviour, consider the nonrelativistic case. Then the charge density can be divided into the density that arises directly from the charge of the boson $\rho^W_e=e\mathbf{\Phi}^*\cdot\mathbf{\Phi}$, density related to its spin $\rho^W_S$ and its quadrupole moment $\rho^W_Q$.
In can be shown \cite{flambaum-kuchiev-prl-07} that the latter two densities together satisfy condition
\begin{equation}
\rho^W_S+\rho^W_Q\,=\,-\mathbf{\nabla}\cdot\mathbf{P}~,
	\label{P}
\end{equation}
where $\mathbf{P}$  is
\begin{equation}
	\mathbf{P}\,=\,-(g-1)\frac{e}{m^2}\,\mathrm{Re}\,
	\big(\,\mathbf{\Phi}^*(\mathbf{\nabla}
	\cdot \mathbf{\Phi})\,\big)~.
	\label{PPhi}
\end{equation}
Here $\mathbf{\Phi}$ is the nonrelativistic wave function of the vector particle,
$g$ represents the conventional $g$-factor, which for W-bosons equals $g=2$.
Clearly, $\mathbf{P}$ in Eq.(\ref{P}) plays a role of polarization. Since the density in Eq.(\ref{P}) equals the divergence of the polarization, it does not contribute to the total charge. Therefore $\rho^W_S+\rho^W_Q$ oscillates, having the ``wrong'' sign in some areas of space. The presence of a derivative in Eq.(\ref{PPhi}) ensures that with increase of energy the contribution of $\rho^W_S+\rho^W_Q$ rises. Thus, one should expect that in the relativistic region the sign of total density can be defined by the 
sign of $\rho^W_S+\rho^W_Q$, which is not determined.

This work was supported by the Australian Research Council.

\end{document}